\documentclass[twocolumn,aip,reprint]{revtex4-2}

\usepackage{amsmath,amssymb,amsfonts,bm}
\usepackage{graphicx,epstopdf}  
\usepackage{color}
\usepackage{extarrows}
\usepackage{enumitem}
\usepackage{empheq}

\usepackage[utf8]{inputenc}
\usepackage[T1]{fontenc}
\usepackage{mathptmx}
\usepackage{etoolbox}

\renewcommand{\d}{{\rm d}}

\newcommand{\ud}{{\mathrm d}}

\newcommand{\w}{\omega}
\newcommand{\sbt}{SBET}
\newcommand{\wti}{\widetilde}
\newcommand{\ti}{\tilde}
\newcommand{\pb}{{\hat\sigma_b}}
\newcommand{\B}{\mbox{\tiny B}}

\newcommand{\ET}{\mbox{\tiny ET}}

\newcommand{\tS}{\mbox{\tiny S}}
\newcommand{\tSS}{\mbox{\tiny SS}}
\newcommand{\tBS}{\mbox{\tiny BS}}
\newcommand{\tSB}{\mbox{\tiny SB}}
\newcommand{\T}{\mbox{\tiny T}}

\newcommand{\SB}{\mbox{\tiny SB}}
\newcommand{\tBB}{\mbox{\tiny BB}}
\newcommand{\dg}{\dagger}
\newcommand{\la}{\langle}
\newcommand{\ra}{\rangle}

\newcommand{\Sec}[1]{Sec.\,\ref{#1}}
\newcommand{\App}[1]{Appendix \ref{#1}}
\newcommand{\nl}{\nonumber \\}
\newcommand{\be}{\begin{equation}}
\newcommand{\ee}{\end{equation}}
\newcommand{\bsube}{\begin{subequations}}
\newcommand{\esube}{\end{subequations}}
\newcommand{\Eq}[1]{Eq.\,(\ref{#1})}
\newcommand{\Eqs}[1]{Eqs.\,(\ref{#1})}
\newcommand{\Fig}[1]{Fig.\,\ref{#1}}

\newcommand{\RN}[1]{%
  \textup{\uppercase\expandafter{\romannumeral#1}}%
}

\newcommand{\vib}{\mbox{\tiny vib}}
\newcommand{\hyb}{\mbox{\tiny hyb}}
\newcommand{\sol}{\mbox{\tiny sol}}
\newcommand{\re}{\mbox{\tiny re}}

\begin{document}
\title{Generalized system-bath entanglement theorem for Gaussian environments}
\author{Yu Su}
\author{Yao Wang}     \email{wy2010@ustc.edu.cn}
\author{Rui-Xue Xu}   \email{rxxu@ustc.edu.cn}
\author{YiJing Yan}
\affiliation{Hefei National Research Center for Physical Sciences at the Microscale,
University of Science and Technology of China, Hefei, Anhui 230026, China}

\date{\today}

\begin{abstract}
A system-bath entanglement theorem (\sbt) with Gaussian environments was
established previously in {\em J.~Chem.~Phys.~}{\bf 152}, 034102 (2020)
in terms of linear response functions.
This theorem connects the system-bath entanglement responses to the local system and bare bath ones.
In this work, we generalize it to correlation functions.
Key steps in derivation are the generalized Langevin dynamics for the hybridizing bath modes as in the previous work,
together with the Bogoliubov transformation mapping the original finite-temperature canonical reservoir
to an effective zero-temperature vacuum via an auxiliary bath.
With the theorem, the system-bath entangled correlations and the bath modes correlations in the full composite space
can be evaluated as long as the bare-bath statistical properties are known
and the reduced system correlations are obtained.
Numerical demonstrations are carried out
for the evaluation of the solvation free energy of an electron transfer system with a certain intramolecular vibrational modes.
\end{abstract}
\maketitle

\section{Introduction}
Many fields of modern research inevitably encounter the system--plus--environments hybridized effects
which can be crucial.\cite{Wei21}
Various developments of quantum dissipation methods are mainly based on Gaussian properties of
environments (thermal baths) being composed of non-interacting harmonic oscillators
coupled linearly to the system.\cite{Cal83587,Gra88115}
Methods cover from perturbative master equation approaches,\cite{Lin76393,Wan53728,Blo571206,   Red651,Yan002068}
stochastic approaches,\cite{Dio97569,Sto994983,Sha045053,Hsi18014103,Hsi18014104,Wan191375,Yan19074106,Che21174111}
the multilayer multiconfiguration time-dependent Hartree method (ML-MCTDH),\cite{Wan031289}
to Feynman--Vernon influence functional path integral\cite{Fey63118}
and its executable propagators,\cite{Mak92435,Mak954600,Mak954611}
together with its differential equivalence,
the hierarchical--equations--of--motion (HEOM)
formalism.\cite{Tan906676,Yan04216,Xu05041103,Tan20020901}

Most of these methods primarily focus on the reduced system dynamics under the influence of thermal environments.
On the other hand, the system-bath entanglement can take essential effect in, for example,
Fano resonances,\cite{Fan611866,Zha09820,Mir102257} vibronic spectroscopies,\cite{Che21244105} and thermal transports.\cite{Wan22044102}
To this end, a dissipaton equation of motion (DEOM) approach has been systematically developed in recent years,
as an exact, quasi-particle theory for system--environment
hybridized dynamics.\cite{Yan14054105,Zha15024112,Xu151816,Wan22170901}
For the dynamics of the reduced system, the DEOM recovers the HEOM.

A system-bath entanglement theorem (\sbt) for Gaussian environments
was established in our previous work.\cite{Du20034102} 
This theorem connects the entangled system-bath linear
response functions in the total system--plus--bath space
to the local responses of associated system variables
and the bath responses of the hybridizing modes in the bare--bath space.
Thus the theorem enables those quantum dissipation methods
which only evaluate the system responses to obtain the system-bath entangled properties as well.

The \sbt\cite{Du20034102} was constructed independent of concrete ensembles.
The key step in establishing it is the generalized Langevin dynamics
for the hybridizing bath modes.
It leads to the \sbt\ existing for any steady state of the composites.
On the other hand, due to the less information of response functions with respect to correlation functions,
it would be inconvenient to further derive the relation between correlations from the original \sbt.\cite{Du20034102}

In this work, we establish a generalized \sbt\ for the connection of correlation functions
among system-bath entangled, local system, and bare bath ones.
It is achieved via the Bogoliubov transformation\cite{Ume95}
mapping the original finite-temperature canonical reservoir
to an effective zero-temperature vacuum by adding an auxiliary bath.
The original \sbt\cite{Du20034102} can be recovered
straightforwardly from the generalized one here.
Like the original one, the generalized \sbt\ here is also established for steady states.
It is shown to satisfy the detailed--balance relation in the condition of the canonical ensemble thermal equilibrium.
The evaluation on expectation values of entangled system-bath properties would just be the special case of correlation functions.

For numerical demonstrations we apply the theorem to evaluate the solvation free energy
of an electron transfer (ET) system involved with several intramolecular vibrational normal modes and embedded
into a Gaussian solvent.
Particularly, based on the generalized SBET,
a multi-scale approach is provided to investigate the solvation effects.
The SBET connects  different dynamical scales in a rigorous way, preserving all the important information such as environmental memory and cross--scale correlations.
As a result, in order to obtain the hybridization free energy for the mixing between solvent and solute molecule which includes both electronic and nuclear degrees of freedom,
we only need to carry out the calculation on the electronic subsystem.
This largely reduces computing cost.
The rest of paper is organized as follows.
The generalized \sbt\ is constructed in \Sec{sbt} with more derivation details given in \App{appa} and
the detailed--balance proof for the canonical ensemble given in \App{appb}.
Numerical demonstrations are presented and discussed in \Sec{num}.
The paper is finally summarized in \Sec{sum}.

\section{Generalized system-bath entanglement theorem}
\label{sbt}
\subsection{Prelude}

Let us start with the total system--plus--bath composite Hamiltonian reading\cite{Du20034102}
\begin{align}\label{HT}
    H_{\T} = H_{\tS} + h_{\B} + H_{\SB}  \equiv
 H_{\tS} + h_{\B} + \sum_u\hat Q_u\hat F_u.
\end{align}
Here, the system Hamiltonian $H_{\tS}$ and dimensionless dissipative modes $\{\hat Q_u\}$ are arbitrary and Hermitian.
The Gaussian environment scenario requires
\begin{align}\label{hB}
    h_{\B} = \sum_j\hbar\omega_{j}\hat a_{j}^\dagger\hat a_{j}
  \quad \text{and}\quad
    \hat F_{u} = \frac{1}{\sqrt{2}}\sum_{j}\hbar c_{uj}(\hat a_{j}^\dagger+\hat a_{j}),
\end{align}
where $\{\hat a^\dg_j\}$
and $\{\hat a_j\}$ are the creation and annihilation operators of bath oscillators.
For the generalized \sbt\ to be established, we consider
 the averages of variables under the following steady state
\begin{align}
    \rho_{\rm st} \equiv \lim_{t_0\to-\infty}e^{iH_{\T}t_0/\hbar}\rho_{\T}(t_0)e^{-iH_{\T}t_0/\hbar},
\end{align}
satisfying $[H_{\T},\rho_{\rm st}]=0$. The total composites are initialized at $t_0$ with
\begin{align}\label{rhoTt0}
    \rho_{\T}(t_0) = \rho_{\tS}(t_0)   \rho_{\B}^{\rm eq}
 \equiv  \rho_{\tS}(t_0) e^{-\beta h_{\B}}/{\rm tr}_{\B}
                                e^{-\beta h_{\B}}.
\end{align}
Here $\beta=1/(k_BT)$
with $k_B$ being the Boltzmann constant and $T$ the temperature.
Correspondingly, denote operators in Heisenberg picture by
\begin{align}\label{Ot}
    \hat O(t) \equiv e^{iH_{\T}(t-t_0)/\hbar}\hat Oe^{-iH_{\T}(t-t_0)/\hbar},
\end{align}
and the correlation functions
\begin{align}\label{corrdef}
    \la\hat A(t)&\hat B(0)\ra\equiv{\rm Tr}\big(e^{iH_{\T}t/\hbar}\hat Ae^{-iH_{\T}t/\hbar}\hat B\rho_{\rm st}\big)\nl
          &=\lim_{t_0\to-\infty}{\rm Tr}\big[e^{iH_{\T}(t-t_0)/\hbar}\hat Ae^{-iH_{\T}(t-t_0)/\hbar}
\nl &\qquad\qquad\quad \times
          e^{-iH_{\T}t_0/\hbar}\hat Be^{iH_{\T}t_0/\hbar}\rho_{\T}(t_0)\big]\nl
          &=\la\hat B(-t)\hat A(0)\ra^\ast,
\end{align}
where $\hat A$ and $\hat B$ are Hermitian operators.
The second equality shows the equivalence between the $\rho_0$ and $\rho_{\rm st}$ descriptions for correlation functions.

The Gaussian bath property is fully characterized via its bare--bath correlation function, namely,
\begin{align}\label{cuvtdef}
    c_{uv}(t) \equiv \la\hat F^{\B}_{u}(t)\hat F^{\B}_{v}(0)\ra_{\B}=c^\ast_{vu}(-t)
\end{align}
where $\hat F^{\B}_{u}(t) \equiv e^{ih_{\B}(t-t_0)/\hbar}\hat F_{u}e^{-i h_{\B}(t-t_0)/\hbar}$
and $\la(\cdots)\ra_{\B} \equiv {\rm tr}_{\B}[(\cdots)\rho^{\rm eq}_{\B}]$
with $\rho^{\rm eq}_{\B}\equiv e^{-\beta h_{\B}}/{\rm Tr}(e^{-\beta h_{\B}})\equiv e^{-\beta h_{\B}}/Z_{\B}$.
The fluctuation--dissipation theorem (FDT) gives\cite{Wei21}
\begin{align}\label{fdt}
    c_{uv}(t) = \frac{\hbar}{\pi}\int_{-\infty}^\infty\!\!\ud\omega\,e^{-i\omega t}\frac{J_{uv}(\omega)}{1-e^{-\beta\hbar\omega}},
\end{align}
with the interacting bath spectral density being
\begin{align}
    J_{uv}(\omega) \equiv \frac{1}{2i}\int_{-\infty}^\infty\!\!\ud t\,e^{i\omega t}\phi_{uv}(t)
\end{align}
and the bare--bath response function
\begin{align}\label{phiuv}
    \phi_{uv}(t) \equiv \frac{i}{\hbar}\la[\hat F^{\B}_{u}(t),\hat F^{\B}_{v}(0)]\ra_{\B}=-\frac{2}{\hbar}{\rm Im}[c_{uv}(t)].
\end{align}
For the Gaussian bath scenario, \Eq{hB}, we can easily obtain
\be\label{phiabsin}
  {\phi}_{uv}(t)=\hbar\sum_j c_{uj} c_{vj} \sin(\w_jt),
\ee
\be\label{Juvw}
J_{uv}(\w)=\frac{\pi}{2}\hbar\sum_j c_{uj} c_{vj}[\delta(\w-\w_j)-\delta(\w+\w_j)],
\ee
and
\be\label{cuvt}
  c_{uv}(t)=  \frac{\hbar^2}{2}\sum_j c_{uj} c_{vj}\left[\bar n_{j}e^{i\w_jt}+(\bar n_{j}+1)e^{-i\w_jt}\right],
\ee
with $\bar n_{j} \equiv (e^{\beta\hbar\omega_{j}}-1)^{-1}$ the average occupation number.
For later use, we define
\be
c^-_{uv}(t)\equiv c_{uv}(t)\quad {\rm and}\quad
c^+_{uv}(t) \equiv [c^-_{uv}(t)]^\ast.
\ee

\subsection{Thermofield decomposition and Langevin dynamics}

For the convenient derivation to obtain the \sbt\ in terms of correlation functions defined in \Eq{corrdef},
we can apply the Bogoliubov transformation\cite{Ume95} by adding an auxiliary bath
$h_{\B}' = -\sum_{j}\hbar\omega_{j}\hat a_{j}^{\prime\dagger}\hat a'_{j}$
to purify $\rho_{\B}^{\rm eq}$ into a vacuum state of the effective bath, $\tilde h_{\B} \equiv  h_{\B} + h_{\B}'$.
Details are given in \App{appa}.
After the Bogoliubov transformation,
let us adopt the thermofield decomposition,\cite{Ume95}
\begin{align}\label{Fdecomp}
    \hat F_{u} = \hat F^+_{u} + \hat F^-_{u},
\end{align}
with
\begin{align}
    \hat F^-_{u} = \frac{1}{\sqrt{2}}\sum_j\hbar{c_{uj}}\big(
     \sqrt{\bar n_{j}}\hat d_{j} + \sqrt{\bar n_{j}+1}\hat d'_{j} \big)
\end{align}
and $\hat F^+_{u} = (\hat F^-_{u})^\dagger$.
Here,
 $       \hat d_{j} \equiv \sqrt{\bar n_{j}+1}\hat a'_{j} - \sqrt{\bar n_{j}}\hat a^\dagger_{j}$ and
 $      \hat d'_{j} \equiv \sqrt{\bar n_{j}+1}\hat a_{j} - \sqrt{\bar n_{j}}\hat a^{\prime\dagger}_{j}$,
as defined in \Eq{apptrans}.

For the effective total Hamiltonian,
$ \wti   H_{\T} =
 H_{\tS} + \ti h_{\B} + \sum_u\hat Q_u\hat F_u$,
the Heisenberg equation of motion gives
$\dot{\hat a}^{\prime}_j(t)=i\omega_j\hat a'_j(t)$, $\dot{\hat a}^{\prime\dagger}_j(t)=-i\omega_j\hat a^{\prime\dagger}_j(t)$, and
\bsube
\begin{align}
\dot{\hat a}^\dg_j(t)&= i\omega_j\hat a_j^\dg(t) +\frac{i}{\sqrt{2}}\sum_v{c_{vj}}\hat Q_v(t),\\
    \dot{\hat a}_j(t)&=-i\omega_j\hat a_j(t)     -\frac{i}{\sqrt{2}}\sum_v{c_{vj}}\hat Q_v(t).
\end{align}
\esube
The formal solutions to the above equations are ${\hat a}^{\prime}_j(t)=e^{i\w_j(t-t_0)}{\hat a}^{\prime}_j$,
${\hat a}^{\prime\dagger}_j(t)=e^{-i\w_j(t-t_0)}{\hat a}^{\prime\dagger}_j$, and
\bsube
\begin{align}
   \hat a^\dg_j(t)&=e^{i\w_j(t-t_0)}\hat a^\dg_j
   +\frac{i}{\sqrt{2}}\sum_v{c_{vj}}
   \int_{t_0}^t\!\!{\rm d}\tau\,e^{i\w_j(t-\tau)}\hat Q_v(\tau),\\
   \hat a_j(t)&=e^{-i\w_j(t-t_0)}\hat a_j
   -\frac{i}{\sqrt{2}}\sum_v{c_{vj}}
   \int_{t_0}^t\!\!{\rm d}\tau\,e^{-i\w_j(t-\tau)}\hat Q_v(\tau).
\end{align}
\esube
Therefore we can obtain
\begin{align}\label{Fpm}
    \hat F_{u}^{\pm}(t) = \hat F^{\pm;\B}_{u}(t) \pm \frac{i}{\hbar}\sum_v\int_{t_0}^t\!\ud\tau\,
    c^{\pm}_{uv}(t-\tau)\hat Q_v(\tau),
\end{align}
with
\be
\hat F^{\pm;\B}_{u}(t) \equiv e^{i\tilde h_{\B}(t-t_0)/\hbar}\hat F^{\pm}_{u}e^{-i\tilde h_{\B}(t-t_0)/\hbar}\,.
\ee
Apparently, $\hat F^+_{u}(t) = [\hat F^-_{u}(t)]^\dagger$.
It is easy to verify that \Eqs{Fdecomp} and (\ref{Fpm}) lead to
\be\label{LangF}
   \hat F_{u}(t)=\hat F^{\B}_{u}(t)-\sum_v\int_{t_0}^t\!\!{\rm d}\tau\,\phi_{uv}(t-\tau)\hat Q_v(\tau).
\ee
This is just the Eq.(7) in Ref.\,\onlinecite{Du20034102},
i.e.\ the equation of Langevin dynamics for the hybridizing bath modes,
which serves as the starting point to the establishment of \sbt.

\subsection{Derivation to the generalized \sbt}

We are now ready to derive the \sbt\ for correlation functions.
By \Eq{apptrace}, the second identity of \Eq{corrdef} can now be recast as
\begin{align}\label{correff}
    \la\hat A(t)\hat B(0)\ra
    =\lim_{t_0\to-\infty}{\wti{\rm Tr}}\big[&
           e^{i\wti H_{\T}(t-t_0)/\hbar}\hat Ae^{-i\wti H_{\T}(t-t_0)/\hbar}\nl &
    \times e^{-i\wti H_{\T}t_0/\hbar}\hat Be^{i\wti H_{\T}t_0/\hbar} \rho_{\tS}(t_0)|\xi\ra\la\xi|\big].
\end{align}
Here, $\wti{\rm Tr}(\cdots)={\rm Tr}[{\rm tr}_{\B}'(\cdots)]$ is the trace over the entire space of $\wti H_{\T}=H_{\T}+h_{\B}'$,
i.e.\ the total composite plus the auxiliary bath. $|\xi\ra$ is denoted as
the vacuum state of the effective bath $\ti h_{\B}$, which is specified in \Eq{vacuum} and proved afterwards.

First of all, the correlations between system operators are 
\begin{align}\label{CSSuv}
    C^{\tSS}_{uv}(t) \equiv \la\hat Q_u(t)\hat Q_v(0)\ra=[C^{\tSS}_{vu}(-t)]^\ast,
\end{align}
which can usually be evaluated by various quantum dissipation methods on the reduced system.
The challenges are the system--bath entangled correlation functions,
\begin{align}\label{CBS}
    C^{\tBS}_{uv}(t) \equiv \la\hat F_u(t)\hat Q_v(0)\ra,
\end{align}
which give also
\be\label{CSBuv}
C^{\tSB}_{uv}(t) \equiv \la\hat Q_u(t)\hat F_v(0)\ra=[C^{\tBS}_{vu}(-t)]^\ast,
\ee
and the bath modes correlations in the full space,
\begin{align}\label{CBB}
    C^{\tBB}_{uv}(t) \equiv \la\hat F_{u}(t)\hat F_{v}(0)\ra.
\end{align}
In the evaluation of \Eq{CBS} or \Eq{CBB}, the difficulty lies in the
$\hat F^{\B}_{u}(t)$--related average in the full space [cf.\,\Eq{LangF}].
This can be overcome by noting that, for any time $t$,
\begin{align}\label{Fxi}
    \hat F^{-;\B}_{u}(t)|\xi\ra\la\xi| =
    |\xi\ra\la\xi|\hat F^{+;\B}_{u}(t) = 0,
\end{align}
and
\begin{align}\label{FQcommu}
    [\hat F^{-;\B}_{u}(t),\hat Q_v] =
    [\hat F^{+;\B}_{u}(t),\hat Q_v] = 0.
\end{align}
Hence the thermofield decomposition, \Eq{Fdecomp}, together with
the corresponding equation of Langevin dynamics, \Eq{Fpm}, are adopted
as the key steps in the derivation of generalized \sbt.

To go on, we separate \Eq{CBS} or \Eq{CBB} into two terms by
\begin{align}\label{sepa}
  \la\hat F_u(t)\hat O(0)\ra&=\la\hat F^+_u(t)\hat O(0)\ra+\la\hat F^-_u(t)\hat O(0)\ra\nl
   &=2{\rm Re}\la\hat F^+_u(t)\hat O(0)\ra+\la[\hat F^-_u(t),\hat O(0)]\ra\,.
\end{align}
The first term in the second identity of \Eq{sepa}
can be evaluated via
\Eq{correff} by substituting \Eq{Fpm} into it with $t_0 \to - \infty$ and using \Eq{Fxi},
while the second term can be evaluated via the first identity of
\Eq{corrdef} then substituting \Eq{Fpm} with $t_0 =0$. Thus we obtain
\bsube
\begin{align}\label{gsbt0a}
C^{\tBS}_{uv}(t)&=-\frac{2}{\hbar}{\rm Im}\bigg[\!\sum_{u'}\int_0^\infty\!\!\ud\tau\,c^+_{uu'}(t+\tau) C_{u'v}^{\tSS}(-\tau)\bigg]\nl
           &\quad -\frac{2}{\hbar}{\rm Im}\!\bigg[\sum_{u'}\int_0^t\!       \ud\tau\,c^+_{uu'}(t-\tau) C_{u'v}^{\tSS}( \tau)\bigg]\nl
           &\quad -\frac{i}{\hbar} \sum_{u'}\int_0^t\!  \ud\tau\,c^-_{uu'}(t-\tau)[C_{u'v}^{\tSS}( \tau) -C_{vu'}^{\tSS}(-\tau)],\\
C^{\tBB}_{uv}(t)&=-\frac{2}{\hbar}{\rm Im}\!\bigg[\sum_{u'}\int_0^\infty\!\!\ud\tau\,c^+_{uu'}(t+\tau) C_{u'v}^{\tSB}(-\tau)\bigg]\nl
           &\quad -\frac{2}{\hbar}{\rm Im}\!\bigg[\sum_{u'}\int_0^t\!       \ud\tau\,c^+_{uu'}(t-\tau) C_{u'v}^{\tSB}( \tau)\bigg]+\la[\hat F^-_u(t),\hat F_v(0)]\ra\nl
           &\quad -\frac{2}{\hbar} \sum_{u'}\int_0^t\!  \ud\tau\,c^-_{uu'}(t-\tau)[C_{u'v}^{\tSB}( \tau) -C_{vu'}^{\tBS}(-\tau)].
\end{align}
\esube
After some elementary algebraic steps, the final expressions can be recast as
\bsube\label{gsbt}
\begin{align}\label{gsbta}
C^{\tBS}_{uv}(t)
           &=      \frac{2}{\hbar}{\rm Im}\bigg[\!\sum_{u'}\int_t^\infty\!\!\ud\tau\,  c_{uu'}(\tau)C_{vu'}^{\tSS}( \tau-t)\bigg]\nl
           &\quad           - \sum_{u'}\int_0^t\!       \ud\tau\,\phi_{uu'}(\tau)C_{u'v}^{\tSS}( t-\tau) ,\\
C^{\tBB}_{uv}(t)
       &=c_{uv}(t)+\frac{2}{\hbar}{\rm Im}\bigg[\!\sum_{u'}\int_t^\infty\!\!\ud\tau\,  c_{uu'}(\tau)C_{vu'}^{\tBS}( \tau-t)\bigg]\nl
           &\quad           - \sum_{u'}\int_0^t\!       \ud\tau\,\phi_{uu'}(\tau)C_{u'v}^{\tSB}( t-\tau) .
\end{align}
\esube
This is the generalized \sbt\ for correlation functions. It can be easily found
to recover the original \sbt\ in Ref.\,\onlinecite{Du20034102}, the Eqs.(12) and (14) there,
in terms of response functions.
As long as the bare--bath statistical properties, for example, the spectral densities
$\{J_{uv}(\omega)\}$ [cf.\,\Eqs{fdt}--(\ref{cuvt})], are known and the reduced system correlations are obtained
via certain quantum dissipation method,
the system-bath entangled correlations and the bath-bath correlations in the full composite space
can be evaluated via \Eq{gsbt}.
Like the original \sbt,\cite{Du20034102} the generalized \sbt\ here is also valid for steady states.
In this way it does not show the fluctuation--dissipation theorem or the detailed--balance relation
satisfied in the condition of the canonical ensemble thermodynamic equilibrium, explicitly.
The proof of the detailed--balance fulfillment is given in \App{appb}.
Finally we note that for the steady state average
${\rm Tr}(\hat A\hat B\rho_{\rm st})=\la\hat A(0)\hat B(0)\ra\equiv\la\hat A\hat B\ra$, \Eq{gsbt} leads to
\bsube\label{stsbt}
\begin{align}\label{stsbta}
\la\hat F_u\hat Q_v\ra
           &=      \frac{2}{\hbar}{\rm Im}\!\sum_{u'}\int_0^\infty\!\!\ud\tau\,   c_{uu'}(\tau)C_{vu'}^{\tSS}( \tau) ,\\
\la\hat F_u\hat F_v\ra
       &=c_{uv}(0)+\frac{2}{\hbar}{\rm Im}\!\sum_{u'}\int_0^\infty\!\!\ud\tau\,   c_{uu'}(\tau)C_{vu'}^{\tBS}( \tau) ,
\label{stsbtb}
\end{align}
\esube
where $\hat F_u$ and $\hat Q_v$ are Hermitian operators.
Note that as long as the system correlations $\{C_{uv}^{\tSS}(t)\}$ are obtained,
the entangled system-bath correlations $\{C_{uv}^{\tBS}(t)\}$ in \Eq{stsbtb} can then be evaluated via
\Eq{gsbt0a} or \Eq{gsbta}.
For any non-Hermitian operator $\hat O$ if involved, we can separate it into
$\hat O=\hat O^{(+)}+i\hat O^{(-)}$ where
\be
\nonumber
  \hat O^{(+)}\equiv(\hat O+\hat O^{\dg})/2\quad {\rm and}\quad
  \hat O^{(-)}\equiv(\hat O-\hat O^{\dg})/(2i),
\ee
to apply the \sbt.

\section{Numerical demonstrations}
\label{num}

Demonstrations of the applications of \Eq{gsbt} to spectroscopic simulations
would be similar to Ref.\,\onlinecite{Du20034102}.
Hence in this work, we consider another application on the evaluation of
thermodynamic properties. Concretely, this will be achieved by using \Eq{gsbt} together with \Eq{stsbt}.
Consider an electron transfer (ET) system, where $|a\ra$ and $|b\ra$ are the two electronic states of the molecule.
The states are coupled to vibrational normal modes.
Electronic states and vibrational modes both interact further with the solvent composed of harmonic oscillators.
We will see in this section  that
we can obtain the solvation free energy for the mixing between solvent and solute molecule which includes both electronic and nuclear degrees of freedom, by only calculating the two--level electronic subsystem dynamics together with using the generalized SBET.
This approach largely reduces computing cost, without any
loss of important informations.

\subsection{Thermodynamic integral formalism}

Consider a system-bath mixing (hybridization) process. As a result of the second law, the Helmholtz free-energy change in
an isotherm process amounts to the reversible work, i.e.
\be\label{DAw}
   A_{\hyb}=\int_{ini}^{final} \delta w_{\rm rev}
\quad {\rm where} \quad \delta w_{\rm rev}={\rm Tr}[(\ud H_{\T})\rho_{\rm eq}],
\ee
and
\be\label{rhoeq}
  \rho_{\rm eq}\equiv e^{-\beta H_{\T}}/{\rm Tr}(e^{-\beta H_{\T}})\equiv e^{-\beta H_{\T}}/Z_{\T}.
\ee
For the simulation on a realistic system, we may recast \Eq{HT} to explicitly include the reorganization term
\begin{align}\label{HTre}
    H_{\T} =  H_{\tS 0}  +  h_{\B} +\sum_u \hat Q_u\hat F_u+\sum_{uv}E^{\re}_{uv}\hat Q_u\hat Q_v,
\end{align}
where $H_{\tS 0}=H_{\tS}-\sum_{uv}E^{\re}_{uv}\hat Q_u\hat Q_v$ and $h_{\B}$ are the isolated system and bath Hamiltonians, respectively.
To perform \Eq{DAw}, let us introduce a hybridization parameter $\lambda$-augmented total Hamiltonian,\cite{Gon20154111,Gon20214115}
\begin{align}\label{lamHTre}
    H_{\T}(\lambda) =  H_{\tS 0}  +  h_{\B} + \lambda\sum_u \hat Q_u\hat F_u+\lambda^2\sum_{uv}E^{\re}_{uv}\hat Q_u\hat Q_v.
\end{align}
The reversible process can then be described with
varying the hybridization parameter $\lambda$ from 0 to 1 gradually.
Note that the reorganization term is of a quadratic order.
For the convenience of implementation,
let us denote
\bsube
\begin{align}
   \la H_{\SB}\ra_{1}&\equiv \sum_u{\rm Tr}[\lambda\hat Q_u\hat F_u\rho_{\rm eq}(\lambda)],\label{HSB1}
 \\  \label{HSB2}
   \la H_{\re}\ra_{2}&\equiv \sum_{uv}{\rm Tr}[\lambda^2E^{\re}_{uv}\hat Q_u\hat Q_v\rho_{\rm eq}(\lambda)].
\end{align}
\esube
With respect to \Eq{DAw}, we have
\begin{align}\label{DAhyb}
   A_{\hyb}=\int_{0}^{1}\!\! 
    \frac{\ud\lambda}{\lambda}\la H_{\SB}\ra_{1} 
                + 2\int_{0}^{1}\!\!
    \frac{\ud\lambda}{\lambda}\la H_{\re}\ra_{2}
 \equiv  A_{\hyb}^{(1)} + A_{\hyb}^{(2)}.
\end{align}
More details and discussions can refer to Refs.\,\onlinecite{Gon20154111} and \onlinecite{Gon20214115}.
It is also easy to find that \Eq{DAhyb} is equivalent to the Kirkwood's thermodynamic integration formalism.\cite{Gon20214115,Kir35300,Zon08041103,Zon08041104}
The $\la H_{\re}\ra_{2}$ in \Eq{HSB2} is the average of reduced system operators,
while the $\la H_{\SB}\ra_{1}$ in \Eq{HSB1} is a system--bath entangled property.
In our previous work,\cite{Gon20154111,Gon20214115}
their evaluations are carried out via the DEOM approach,
which is exact but would be time-consuming for large systems.
Now we can apply \sbt\    
to calculate them more efficiently.
Numerical results in the FIG.\,1 of Ref.\,\onlinecite{Gon20154111} have been repeated.
In the next subsection, we demonstrate the solvation free energy evaluation for an electron-transfer (ET) system
with a certain intramolecular vibrational modes.
We will see that not only \Eq{stsbt} but also \Eq{gsbt} play the key roles during the evaluation.

\begin{figure}[h]
\includegraphics[width=0.5\textwidth]{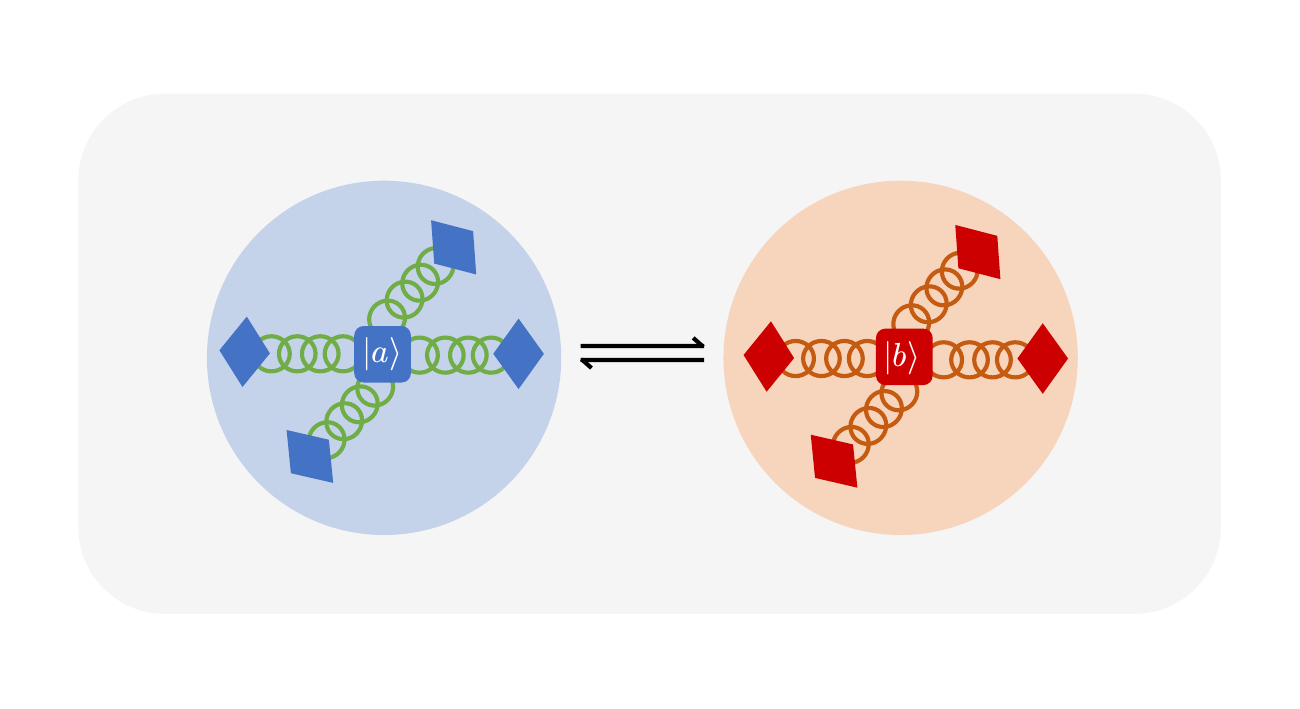}
\caption{
Schematic diagram of the model for numerical demonstrations. The $|a\ra$ and $|b\ra$ are the two electronic states of the molecule. Included are also the  vibrational normal modes. They all interact with the solvent composed of harmonic oscillators.   When the state transits between $|a\ra$ and $|b\ra$, the normal modes and the solvent oscillators are all displaced and reorganized.
}\label{fig1}
\end{figure}

\subsection{Model of ET system}

We set $\hbar=1$ in the following part of this section.
Consider an ET system with the total Hamiltonian being
\be\label{Het}
  H_{\T}=H_a|a\ra\la a|+(H_b+E^\circ)|b\ra\la b|+V(|a\ra\la b|+|b\ra\la a|).
\ee
Here, $E^\circ$ is the reaction
endothermicity and $V$ is the transfer coupling strength.
$H_a$ and $H_b$ are the nuclei--solvent Hamiltonians for the ET system
in the donor and acceptor states, respectively.
They are assumed of the Caldeira--Leggett's model,\cite{Wei21,Cal83587} i.e.
\begin{align}
\label{Henv}
 H_a
&=\sum_{n=1}^{N_{\vib}}\frac{\Omega_{n}}{2}(\hat p_n^2+\hat q_n^2)
+\sum_{k=1}^{N_{\sol}}\frac{\w_k}{2}
 \Bigg\{\ti p^2_k+\bigg[\ti x_k-\bigg(\sum_{n=1}^{N_{\vib}}\frac{c_{nk}}{\w_k}\hat q_n\bigg)\bigg]^2\Bigg\},
\end{align}
and $H_b$ is similar to $H_a$ but with linearly displaced
$\{q'_n=q_n-D_n\}$ and $\{x'_k=x_k-d_k\}$.
The first term in \Eq{Henv} is the Hamiltonian of the involved intramolecular vibrational normal modes,
denoted as $H_{\vib}$ afterwards.
The solvent Hamiltonian is
\be
h_{\sol}=\frac{1}{2}\sum_{k=1}^{N_{\sol}}\w_k(\ti p^2_k+\ti x^2_k).
\ee
$N_{\vib}$ and $N_{\sol}$ are the numbers of the degrees of freedom of
the intramolecular normal modes and the solvent, respectively.
See the schematic diagram of this model in \Fig{fig1}.

Before going on, let us introduce the solvent force operators
\be
\hat X_{n}=\sum_k c_{nk}\ti x_k
\quad {\rm and}
\quad
\hat Y=\sum_k\omega_k d_k\ti x_k.
\ee
Their participation in the total Hamiltonian [\Eq{Het}] will be seen soon later.
Denote
 the involving solvent response functions as
\bsube\label{solphi}
\begin{align}\label{phixx}
  \varphi^{xx}_{mn}(t)&\equiv i\la[\hat X^{\sol}_{m}(t),\hat X_{n}]\ra_{\sol}
    =\sum_kc_{mk}c_{nk}\sin(\w_kt)\nl &=i\la[\hat X^{\sol}_{n}(t),\hat X_{m}]\ra_{\sol}=\varphi^{xx}_{nm}(t),\\
  \varphi^{xy}_n(t)&\equiv i\la[\hat X^{\sol}_{n}(t),\hat Y]\ra_{\sol}
    =\sum_k\w_{k}d_kc_{nk}\sin(\w_kt)\nl &=i\la[\hat Y^{\sol}(t),\hat X_{n}]\ra_{\sol},\\
  \varphi^{yy}(t)&\equiv i\la[\hat Y^{\sol}(t),\hat Y]\ra_{\sol}=\sum_k\w^2_{k}d^2_k\sin(\w_kt),
\end{align}
\esube
and for any function $f(t)$ the frequency resolution,
 $  \ti f(\w)\equiv\int^{\infty}_{0}\!\!\d t\, e^{i\w t} f(t)$.  
Here, $\hat O^{\sol}(t)=e^{ih_{\sol}t}\hat Oe^{-ih_{\sol}t}$ and
 $\la\cdots\ra_{\sol}= {\rm tr}_{\sol}(\cdots e^{-\beta h_{\sol}})
/{\rm tr}_{\sol}e^{-\beta h_{\sol}}$.

   Note that $|a\ra\la a|+|b\ra\la b|=1$ and the Huang–Rhys factor $S_n=D_n^2/2$.
 Denote also $\pb\equiv|b\ra\la b|$
and $\hat V_{\ET}\equiv V(|a\ra\la b|+|b\ra\la a|)$.
The total Hamiltonian of \Eq{Het} can be recast
in terms of \Eq{HTre} with $h_{\B}=h_{\sol}$,
\begin{align}\label{ETHS0}
  H_{\tS 0}=H_{\vib}+\hat V_{\ET}+\Big[E^\circ+
  \sum_{n=1}^{N_{\vib}}{\Omega_n}(S_n-D_n\hat q_n)\Big]\pb,
\end{align}
the multi-dissipative-mode system-bath interaction term,
\begin{align}\label{ETQF}
   \sum_u\hat Q_u\hat F_u=
     \pb \bigg(\sum_{n=1}^{N_{\vib}} D_n \hat X_n-\hat Y\bigg)
     -\sum_{n=1}^{N_{\vib}}\hat q_n\hat X_n\,,
\end{align}
and the multi-mode reorganization term,
\begin{align}\label{ETreQQall}
\sum_{uv}E^{\re}_{uv}\hat Q_u\hat Q_v=
  E^{\re}_{bb}\pb+\sum_n E^{\re}_{nb}\hat q_n\pb
  +\sum_{mn}E^{\re}_{mn}\hat q_m\hat q_n\,,
\end{align}
where
\be
\begin{split}
 E^{\re}_{bb}&=
  \frac{1}{2}\Big[\sum_{mn}\ti\varphi^{xx}_{mn}(0)D_mD_n  -2\sum_n \ti\varphi^{xy}_n(0)D_n
        +\ti\varphi^{yy}(0) \Big],
\\
 E^{\re}_{nb}&=\ti\varphi^{xy}_n(0)
 -\sum_{m}\ti\varphi^{xx}_{mn}(0)D_m\,,
\qquad
     E^{\re}_{mn}=\frac{1}{2}\ti\varphi^{xx}_{mn}(0).
\end{split}
\ee
Thus in the calculation of $A_{\hyb}$ via \Eq{DAhyb}
for the process of the ET molecule embedded into the solvent,
the involved intermediate quantities in the integrands of \Eq{DAhyb} are
$\{\la\hat X_n\pb\ra_{1}\}$,
$  \la\hat Y  \pb\ra_{1}$,
$\{\la \hat X_n\hat q_n\ra_{1}\}$,
$\la\pb\ra_{2}$,
$\{\la\hat q_n\pb\ra_{2}\}$,
$\{\la\hat q_m\hat q_n\ra_{2}\}$.
Here $\la\cdots\ra_1$ and $\la\cdots\ra_2$ are similar to \Eq{HSB1} and \Eq{HSB2}, respectively.
To obtain them, the DEOM approach\cite{Gon20154111,Gon20214115} for the above model [\Eqs{ETHS0}--(\ref{ETreQQall})]
will be very time-consuming.
This can be overcome via the \sbt\ developed in \Sec{sbt} as below.

\subsection{Model to two-state system}

Equation (\ref{Het}) can also be recast as
\be\label{totalEre}
H_{\T}=\Big(E^\circ+E^{\re}_{bb}+\sum_n\Omega_nS_n\Big)\pb+\hat V_{\ET}+\pb\hat F+H_a\,.
\ee
The induced overall vibration-plus-solvent force to the two--electronic--state system is
\begin{align}\label{hatF12}
    \hat F = \sum_nD_n\hat X_{n}-\hat Y +\sum_n \big(E^{\re}_{nb}-\Omega_nD_n\big) \hat q_n\,.
\end{align}
The overall force-force response function is
\be\label{allPhi}
\Phi(t)\equiv i\la [\hat F^{(a)}(t), \hat F]\ra_a\equiv i\la [e^{iH_at}\hat Fe^{-iH_at}, \hat F]\ra_a \,,
\ee
where $\la\cdots\ra_a
  \equiv {\rm Tr}_a(\cdots e^{-\beta H_a})
/{\rm Tr}_ae^{-\beta H_a}$.
Equations (\ref{totalEre})--(\ref{hatF12}) constitute the multi-vibrational-mode generalization of the single-mode exciton system
in Ref.\,\onlinecite{Che21244105}.
The overall response function can be obtained
following the similar derivation there,\cite{Che21244105}
via its frequency resolution, $\wti\Phi(\w)$.
Assume the solvent effects on different vibrational normal modes are un-correlated,
i.e.\,$\varphi^{xx}_{mn}(t)=\delta_{mn}\varphi^{xx}_{nn}(t)$,
leading to also $E^{\re}_{nb}=\ti\varphi^{xy}_n(0)
 -\ti\varphi^{xx}_{nn}(0)D_n$.
Denote $\bar D_n\equiv E^{\re}_{nb}-\Omega_nD_n$.
We can obtain\cite{Che21244105,Du20034102}
\begin{align}
   \wti\Phi(\w)&=\sum_n[D_n^2\ti\varphi^{xx}_{nn}(\w)-2D_n\ti\varphi^{xy}_{n}(\w)]+\ti\varphi^{yy}(\w)
\nl &\quad +\sum_n[\bar D_n+D_n\ti\varphi^{xx}_{nn}(\w)-\ti\varphi^{xy}_{n}(\w)]^2\ti\phi^{qq}_{nn}(\w),
\end{align}
where\cite{Che21244105,Wei21}
\be\label{chiqqn}
\ti\phi^{qq}_{nn}(\w)=
\frac{\Omega_n}{\Omega_n^{2}-\w^2-\Omega_n[\ti\varphi^{xx}_{nn}(\w)-\ti\varphi^{xx}_{nn}(0)]}\,.
\ee

\begin{figure}[h]
\includegraphics[width=0.45\textwidth]{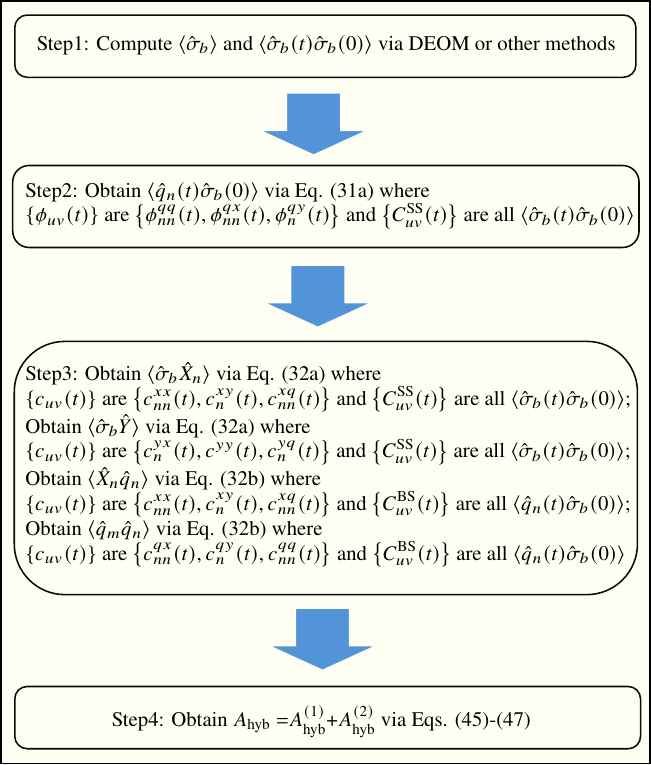}
\caption{
Flowchart for the calculation of the hybridization free energy.
Step 1 only involves the calculation on a two-level system.
In Steps 2 and 3, we first obtain $\phi_{nn}^{qx}(t)$, $\phi_{n}^{qy}(t)$, $\phi_{nn}^{xx}(t)$, $\phi_{n}^{xy}(t)$, and $\phi^{yy}(t)$ via \sbt\ in Ref.\,\onlinecite{Du20034102}.
The first two use the Eq.\,(17) and the last three use the Eq.\,(18) of  Ref.\,\onlinecite{Du20034102}.
The corresponding $c_{nn}^{qx}(t)$, $c_{n}^{qy}(t)$, $c_{nn}^{xx}(t)$, $c_{n}^{xy}(t)$, and $c^{yy}(t)$ are obtained via the FDT, \Eq{fdt}.
}\label{fig2}
\end{figure}

Turn to the solvation free energy $A_{\hyb}$ evaluation.
As just mentioned, the key quantities to be calculated are
$\{\la\hat X_n\pb\ra_{1}\}$,
$  \la\hat Y  \pb\ra_{1}$,
$\{\la \hat X_n\hat q_n\ra_{1}\}$,
$\la\pb\ra_{2}$,
$\{\la\hat q_n\pb\ra_{2}\}$,
$\{\la\hat q_m\hat q_n\ra_{2}\}$.
Given conditions are the solvent response functions [\Eq{solphi}]
in terms of their frequency-domain resolutions,
from which $\{\ti\phi^{qq}_{nn}(\w)\}$ and $\wti\Phi(\w)$ are also determined.
For each selected $\lambda$, the flowchart is shown in \Fig{fig2}.
The procedure repeats for $\lambda$ varying from 0 to 1
until the integrals in \Eq{DAhyb} are converged.

\begin{figure}[h]
\includegraphics[width=0.4\textwidth]{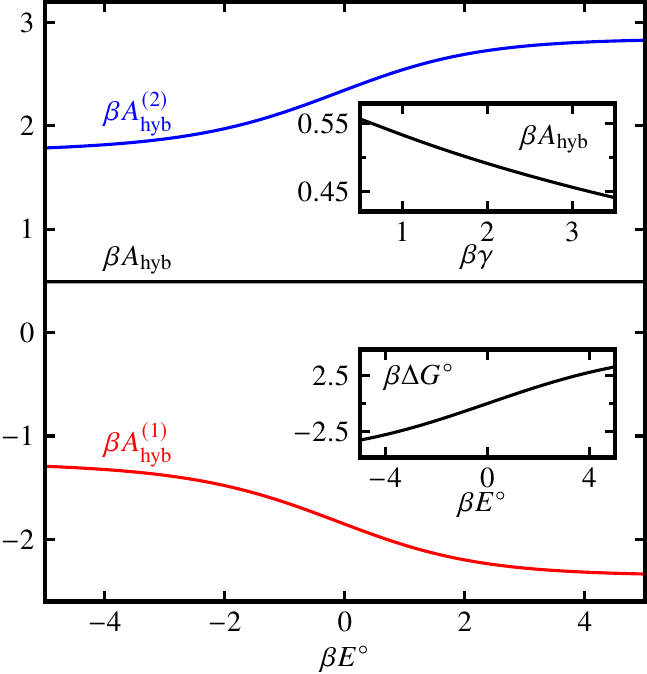}
\caption{
The hybridization free energy $A_{ \hyb}$ (in black), together with $A_{ \hyb}^{(1)}$ (in red) and $A_{ \hyb}^{(2)}$ (in blue), with respective to $E^{\circ}$. The temperature is $300$K.
The main panel adopts $\beta\gamma=2$. See other parameters in the main text.
The hybridization free energy (black line) changes  very little with  $E^{\circ}$, but more explicitly with $\gamma$ (upper inset).
Depicted in the lower inset is $\beta \Delta G^{\circ}=-\ln K$ with $K$ being the  equilibrium constant.
}\label{fig3}
\end{figure}

Figure \ref{fig3} exhibits the obtained hybridization free energy $A_{\hyb}$ with different values of the reaction endothermicity $E^{\circ}$.
We choose $N_{\vib}=2$.
The temperature is $300$K.
In the demonstration, we adopt
\be
\begin{split}
\ti \varphi^{xx}_{nn}(\w)=\frac{2\eta_{n}^{xx}\gamma}{\gamma-i\w},
\\
\ti \varphi^{xy}_{n}(\w)=\frac{2\eta_{n}^{xy}\gamma}{\gamma-i\w},
\\
\ti \varphi^{yy}_{}(\w)=\frac{2\eta^{yy}\gamma}{\gamma-i\w}.
\end{split}
\ee
Parameters for the main panel of \Fig{fig3} are selected as
$\beta V=1$, $\beta \Omega_1=4$, $\beta \Omega_2=1$, $D_1=D_2=0.5$, and $\beta\gamma=\beta\eta^{xx}_{11}=\beta\eta^{xx}_{22}=2$.
We choose $\eta_{n}^{xy}=D_n\eta_{nn}^{xx}$ and $\eta^{yy}=\sum_n D_n^2 \eta_{nn}^{xx}$
which constitutes the Brownian vibration condition.\cite{Che21244105}
In \Fig{fig3}, it is observed that the total $A_{\hyb}$ 
changes little with $E^{\circ}$, but changes more explicitly with the solvent friction $\gamma$ (upper inset).
Depicted in the lower inset is $\beta \Delta G^{\circ}=-\ln K$ with $K$ being the  equilibrium constant obtained from the computed equilibrium populations.
The contribution of $A_{\hyb}^{(1)}$ due to the solvent--solute interaction  is negative, while that of $A_{\hyb}^{(2)}$ due to the solvent reorganization is positive.
They altogether amount to a positive $A_{\hyb}$ which implies that external work  is necessary for the reversible solute--solvent mixing.
The absolute values of $A^{(1)}_{\hyb}$ and $A^{(2)}_{\hyb}$ share a similar behaviour with  $\Delta G^{\circ}$,  indicating their relevance with the extent of reaction.
Note in our model for the molecule at the electronic state $a$ or $b$,
its interaction with the solvent is actually of the same effect.
Thus $A_{\rm hyb}$ is nearly unchanged in the reaction system
$a \rightleftharpoons b$ with different system
endothermicity $E^{\circ}$.

\section{Summary}
\label{sum}

To summarize, this work generalizes the system--bath entanglement theorem (SBET), previously established for response functions,\cite{Du20034102} to correlation functions.
The derivation involves the use of generalized Langevin dynamics for the hybridizing bath modes and the Bogoliubov transformation.
The latter maps the finite--temperature canonical reservoir to an effective zero--temperature vacuum.
As a result, the generalized SBET
connects the system--bath entanglement correlations to the local system and bare bath ones.
It facilitates the evaluation of system--bath entangled correlations and bath mode correlations in the full composite space.

The demonstrations are carried out on computing the solvation free energy of an electron transfer system with specific intramolecular vibrational modes,  exhibiting the practical utility of the generalized SBET.
Particularly, based on the SBET,
we develop a multi-scale approach to investigate the solvation effects,  by only computing  the electronic subsystem.
In this approach, we separate different dynamical scales into electronic, vibrational and solvent parts.
The SBET connects different scales rigorously and largely reduces computing cost, without any
loss of important information such as environmental memory and cross--scale correlations.
This provides an approach to investigate the large scale effects by only computing the small center system.

\begin{acknowledgements}
	Support from the Ministry of Science and Technology of China (Grant No.\ 2021YFA1200103) and
the National Natural Science Foundation of China (Grant Nos.\ 22103073 and 22173088)
is gratefully acknowledged.
\end{acknowledgements}

\appendix
\section{Bogoliubov transformation}
\label{appa}
In order to perform the Bogoliubov transformation,\cite{Ume95}
we shall first double the bath degrees of freedom by adopting an auxiliary bath as
\begin{align}
    h_{\B}' = -\sum_{j}\hbar\omega_{j}\hat a_{j}^{\prime\dagger}\hat a'_{j}.
\end{align}
We will see later that the effective bath Hamiltonian [cf.\,\Eq{hB}]
\be
\tilde h_{\B} \equiv  h_{\B} + h_{\B}'
\ee
retain the same form as
$h_{\B}+h_{\B}'=\sum_{j}\hbar\omega_{j}(\hat a_{j}^\dagger\hat a_{j}-\hat a_{j}^{\prime\dagger}\hat a'_{j})$
after the transformation [\Eq{eqA6}].
Define the Bogoliubov transformation,
\begin{subequations}\label{apptrans}
    \begin{align}
        \hat d_{j} &\equiv \sqrt{\bar n_{j}+1}\hat a'_{j} - \sqrt{\bar n_{j}}\hat a^\dagger_{j},\\
        \hat d'_{j} &\equiv \sqrt{\bar n_{j}+1}\hat a_{j} - \sqrt{\bar n_{j}}\hat a^{\prime\dagger}_{j}.
    \end{align}
\end{subequations}
The inverse transformation reads
\begin{subequations}
    \begin{align}
        \hat a_{j} &= \sqrt{\bar n_{j}+1}\hat d'_{j} + \sqrt{\bar n_{j}} \hat d_{j}^\dagger, \\
        \hat a'_{j} &= \sqrt{\bar n_{j}+1}\hat d_{j}  + \sqrt{\bar n_{j}} \hat d_{j}^{\prime\dagger}.
    \end{align}
\end{subequations}
It is easy to verify the transformation conserve the canonical commutators,
\be
\begin{split}
    &[\hat d'_{j},\hat d'_{k}] =[\hat d'_{j},\hat d_{k}]
      =[\hat d'_{j},\hat d^\dg_{k}] = [\hat d_{j},\hat d_{k}] = 0,\nl
    &[\hat d'_{j},\hat d_{k}^{\prime\dagger}] = [\hat d_{j},\hat d_{k}^\dagger] = \delta_{jk}.
\end{split}
\ee
After the Bogoliubov transformation, we have that
\begin{align}\label{eqA6}
    h_{\B} + h_{\B}' = \sum_{j}\hbar\omega_{j}(\hat d^{\prime\dagger}_{j}\hat d'_{j}-\hat d^\dagger_{j}\hat d_{j}).
\end{align}

Now consider the state
\begin{align}\label{vacuum}
    |\xi\ra \equiv \prod_{j}|\xi_{j}\ra
\
 \
 {\rm with}
\  \
    |\xi_{j}\ra = \frac{1}{\sqrt{Z_{j}}}\sum_{n_{j}=0}^\infty e^{-n_j\beta\hbar\omega_{j}/2}|n_{j}\ra|n_{j}\ra'.
\end{align}
Here, we introduce the phonon's number states
$|n_{j}\ra  \equiv \frac{1}{\sqrt{n_{j}!}}(\hat a^\dagger_{j})^{n_{j}}|{\bf 0}\ra_{\B}$ and
$|n_{j}\ra' \equiv \frac{1}{\sqrt{n_{j}!}}(\hat a^{\prime\dagger}_{j})^{n_{j}}|{\bf 0}\ra_{\B}'$,
for the original bath $h_{\B}$ and the auxiliary bath $h_{\B}'$, respectively.
$Z_{j}\equiv (1-e^{-\beta\hbar\omega_{j}})^{-1}$ is the partition function for the harmonic mode of frequency $\omega_{j}$.
It is easy to see that for the $\rho_{\B}^{\rm eq}=e^{-\beta h_{\B}}/{\rm tr}_{\B}e^{-\beta h_{\B}}$ as in \Eq{rhoTt0},
there is
\be\label{apptrace}
  \rho_{\B}^{\rm eq}={\rm tr}_{\B}'\left(|\xi\ra\la\xi|\right),
\ee
where ${\rm tr}'_{\B}$ is the partial trace over the auxiliary bath space.
Furthermore we have
\begin{align*}
   \sqrt{Z_{j}}\hat d'_{j}|\xi_{j} \rangle
    =&\sum_{n_{j}=0}^{\infty}\!\!e^{-n_j\beta\hbar\omega_{j}/2}\big(\sqrt{\bar{n}_{j}+1}\hat a_{j}
    -\sqrt{\bar{n}_{j}}\hat a_{j}^{\prime\dagger}\big)|n_{j}\rangle |n_{j}\rangle'\\
    =&
    \sum_{n_{j}=1}^{\infty}e^{-n_j\beta\hbar\omega_{j}/2}\sqrt{\bar{n}_{j}+1}\sqrt{n_{j}}|n_{j}-1\rangle |n_{j}\rangle'\\
    &-\sum_{n_{j}=0}^{\infty}e^{-n_j\beta\hbar\omega_{j}/2}\sqrt{\bar{n}_{j}}\sqrt{n_{j}+1}|n_{j}\rangle |n_{j}+1\rangle',\\
    =&
    \sum_{n_{j}=0}^{\infty}e^{-(n_{j}+1)\beta\hbar\omega_{j}/2}\sqrt{1+\bar{n}_{j}}\sqrt{n_{j}+1}|n_{j}\rangle |n_{j}+1\rangle'\\
    &-\sum_{n_{j}=0}^{\infty}e^{-n_j\beta\hbar\omega_{j}/2}\sqrt{\bar{n}_{j}}\sqrt{n_{j}+1}|n_{j}\rangle |n_{j}+1\rangle',\\
    =&
    \sum_{n_{j}=0}^{\infty}\!\!e^{-n_j\beta\hbar\omega_{j}/2}\big(e^{-\beta\hbar\omega_{j}/2}\sqrt{\bar{n}_{j}+1}-\sqrt{\bar{n}_{j}}\big)\\
    &\times\sqrt{n_{j}+1}|n_{j}\rangle |n_{j}+1\rangle'=0.
\end{align*}
In the last step, we have used the relation
$e^{-\beta\hbar\omega_{j}}={\bar{n}_{j}}
                   /(\bar{n}_{j}+1)$.
Similarly, we can obtain also $\hat d_{j}|\xi_{j} \ra =0$.
It is thus proved that the state, $|\xi\ra$, is actually the vacuum state of the effective bath $\tilde h_{\B}$.

\section{Detailed--balance relation}
\label{appb}

In the canonical thermodynamic equilibrium condition, there is
\begin{align}\label{appb1}
    \la\hat A(t)&\hat B(0)\ra_{\rm eq}
    ={\rm Tr}\big(e^{iH_{\T}t/\hbar}\hat A
    e^{-iH_{\T}t/\hbar}\hat B\rho_{\rm eq}\big),
\end{align}
with $\rho_{\rm eq}$ defined in \Eq{rhoeq}.
The detailed--balance relation reads
\be\label{dbr}
    \la\hat A(t)\hat B(0)\ra^\ast_{\rm eq}=
    \la\hat A(t-i\beta\hbar)\hat B(0)\ra_{\rm eq}\,.
\ee
To prove \Eq{gsbt} satisfy \Eq{dbr} in the condition of $\rho_{\rm st}=\rho_{\rm eq}$,
we recast the main terms involved in \Eq{gsbt} in a unified form as
[cf.\,\Eqs{phiuv}, (\ref{CSSuv}) and (\ref{CSBuv}) together with some variables' changes of the involved integrals]
\begin{align}\label{calC}
{\cal C}(t)
           &\equiv
    {\rm Im}\bigg[\int_0^\infty\!\!\ud\tau\,
      c_{uu'}(t+\tau)C_{OO'}(\tau)\bigg]\nl
           &\quad  +\int_0^t\!\ud\tau\,
           {\rm Im}[c_{uu'}(t-\tau)]C_{O'O}(\tau).
\end{align}
Here, $\hat O$ and $\hat O'$ are arbitrary Hermitian operators of the system or bath.
We need to prove ${\cal C}^{\ast}(t)={\cal C}(t-i\beta\hbar)$.
To do that we further expand \Eq{calC} into four terms as
\begin{align}\label{calCexpa}
{\cal C}(t)
           &=
    \frac{i}{2}\int_0^\infty\!\!\ud\tau\,
      c^\ast_{uu'}(t+\tau)C^\ast_{OO'}(\tau)
      \nl
           &\quad
                    -\frac{i}{2}\int_0^\infty\!\!\ud\tau\,
      c_{uu'}(t+\tau)C_{OO'}(\tau)
      \nl
           &\quad  +\frac{i}{2}\int_0^t\!\ud\tau\,
           c^\ast_{uu'}(t-\tau)C_{O'O}(\tau)
      \nl
           &\quad
                      -\frac{i}{2}\int_0^t\!\ud\tau\,
           c_{uu'}(t-\tau)C_{O'O}(\tau)
      \nl
           &\equiv \frac{\hbar}{2}\big[{\cal C}_1(t)+{\cal C}_2(t)+{\cal C}_3(t)+{\cal C}_4(t)\big].
\end{align}

Let us denote the eigenstate and eigenenergy of the bath Hamiltonian $h_{\B}$ as
$|m\ra$ and $\varepsilon_m$, i.e.\ $h_{\B}|m\ra=\varepsilon_m|m\ra$.
Denote the eigenstate and eigenenergy of the total Hamiltonian $H_{\T}$ as
$|\alpha\ra$ and $E_\alpha$, i.e.\ $H_{\T}|\alpha\ra=E_\alpha|\alpha\ra$.
We can then recast $c_{uu'}(t)$ [\Eq{cuvtdef}] and $C_{OO'}(t)$ [\Eq{appb1} or \Eq{corrdef} with $\rho_{\rm st}=\rho_{\rm eq}$]
as the following expressions, respectively,
\begin{align}\label{appb6}
    c_{uu'}(t)&= \frac{1}{Z_{\B}}\sum_{m,m'} e^{i(\varepsilon_m-\varepsilon_{m'})t/\hbar}
       e^{-\beta\varepsilon_m}F_{u;mm'}F_{u';m'm}\,,
\end{align}
and
\begin{align}\label{appb7}
    C_{OO'}(t) = \frac{1}{Z_{\T}}\sum_{\alpha,\alpha'}e^{i(E_{\alpha}-E_{\alpha'})t/\hbar}e^{-\beta E_\alpha}
    O_{\alpha\alpha'}O'_{\alpha'\alpha}\,.
\end{align}
Substituting
into \Eq{calCexpa}, we obtain
\begin{align}
{\cal C}_1(t)    &=\frac{1}{Z_{\B}Z_{\T}}\sum_{m,m'}\sum_{\alpha,\alpha'}
    e^{i(\varepsilon_{m}-\varepsilon_{m'})t/\hbar}  e^{-\beta(\varepsilon_{m'}+E_{\alpha'})}\nl
&\quad\times    F_{u;mm'}F_{u';m'm}O_{\alpha\alpha'}
                                  O'_{\alpha'\alpha}\nl
    &\quad\times\bigg[-\frac{1}{(\varepsilon_m-\varepsilon_{m'}) + (E_\alpha - E_{\alpha'})}
    \nl
    &\qquad\ \ \,  + i\pi\delta(\varepsilon_m-\varepsilon_{m'} + E_\alpha - E_{\alpha'}) \bigg],\nl
{\cal C}_2(t)    &=\frac{1}{Z_{\B}Z_{\T}}\sum_{m,m'}\sum_{\alpha,\alpha'}e^{i(\varepsilon_{m}-\varepsilon_{m'})t/\hbar}
e^{-\beta(\varepsilon_m+E_{\alpha})}\nl
  &\quad\times   F_{u;mm'}F_{u';m'm}O_{\alpha\alpha'}
                                   O'_{\alpha'\alpha}\nl
    &\quad\times\bigg[ \frac{1}{(\varepsilon_m-\varepsilon_{m'}) + (E_\alpha - E_{\alpha'})}
\nl
    &\qquad\ \ \,      - i\pi\delta(\varepsilon_m-\varepsilon_{m'} + E_\alpha - E_{\alpha'}) \bigg],\nl
{\cal C}_3(t)    &=
\frac{1}{Z_{\B}Z_{\T}}\sum_{m,m'}\sum_{\alpha,\alpha'} e^{-\beta(\varepsilon_{m'}+E_{\alpha'})}
    F_{u;mm'}F_{u';m'm} O_{\alpha\alpha'}
                       O'_{\alpha'\alpha}\nl
    &\quad\times \frac{e^{i(\varepsilon_m-\varepsilon_{m'})t/\hbar}-e^{-i(E_\alpha-E_{\alpha'})t/\hbar}}
    {(\varepsilon_m-\varepsilon_{m'}) + (E_\alpha-E_{\alpha'})}\,,\nl
{\cal C}_4(t)
    &=\frac{1}{Z_{\B}Z_{\T}}\sum_{m,m'}\sum_{\alpha,\alpha'} e^{-\beta(\varepsilon_m+E_{\alpha'})}
    F_{u;mm'}F_{u';m'm} O_{\alpha\alpha'}
                       O'_{\alpha'\alpha}\nl
    &\quad\times \frac{e^{-i(E_\alpha-E_{\alpha'})t/\hbar}-e^{i(\varepsilon_m-\varepsilon_{m'})t/\hbar}}
    {(\varepsilon_m-\varepsilon_{m'}) + (E_\alpha-E_{\alpha'})}\,.\nonumber
\end{align}
Altogether, we have
\begin{align}\label{appb8}
{\cal C}(t)
    &=\frac{\hbar}{2Z_{\B}Z_{\T}}\sum_{m,m'}\sum_{\alpha,\alpha'}
       \frac{F_{u;mm'}F_{u';m'm} O_{\alpha\alpha'}O'_{\alpha'\alpha}}
       {(\varepsilon_m-\varepsilon_{m'}) + (E_\alpha - E_{\alpha'})} \nl
&\quad\times
    \Big\{      e^{-i(E_\alpha-E_{\alpha'})t/\hbar}
\big[{e^{-\beta(\varepsilon_m+E_{\alpha'})}- e^{-\beta(\varepsilon_{m'}+E_{\alpha'})}}\big]\nl
    &\qquad\   +       e^{i(\varepsilon_{m}-\varepsilon_{m'})t/\hbar}
\big[e^{-\beta(\varepsilon_m+E_{\alpha})}-e^{-\beta(\varepsilon_{m}+E_{\alpha'})}\big]\Big\}\,.
\end{align}
We can thus easily find ${\cal C}^{\ast}(t)={\cal C}(t-i\beta\hbar)$ by \Eq{appb8}.
It is then straightforward to finish the proof that the \sbt\ [\Eq{gsbt}], established in general for steady states,
satisfies the detailed--balance relation at the condition of the thermal canonical equilibrium.


\end{document}